\documentclass{ws-mpla}

\def\be{\begin{equation}}
\def\ee{\end{equation}}
\def\ba{\begin{eqnarray}}
\def\ea{\end{eqnarray}}
\newcommand{\fr}[2]{\frac{#1}{#2}}

\begin{document}

\markboth{Seokcheon Lee} {Palatini $f(R)$ Cosmology (Paper's
Title)}

\catchline{}{}{}{}{}

\title{PALATINI $f(R)$ COSMOLOGY}

\author{\footnotesize SEOKCHEON LEE \footnote{Institute of Physics,
Academia Sinica, Taipei, Taiwan 11529, R.O.C.}}

\address{Institute of Physics, Academia Sinica, \\
Taipei, Nankang, 11529, Republic of China. \\
skylee@phys.sinica.edu.tw}

\maketitle

\pub{Received (Day Month Year)}{Revised (Day Month Year)}

\begin{abstract}
We investigate the modified gravity theories in terms of the
effective dark energy models. We compare the cosmic expansion
history and the linear growth in different models. We also study
the evolution of linear cosmological perturbations in modified
theories of gravity assuming the Palatini formalism. We find the
stability of the superhorizon metric evolution depends on models.
We also study the matter density fluctuation in the general gauge
and show the differential equations in super and sub-horizon
scales.

\keywords{Modified Gravity; Palatini Formalism.}
\end{abstract}

\ccode{PACS Nos.: include PACS Nos.}

\section{Introduction}

Type Ia supernova distance-redshift measurement shows that the
present expansion of the universe is accelerating \cite{SNe}. To
explain this phenomena we need to introduce a homogeneous
component of energy with a negative pressure, dubbed dark energy
\cite{DE}, or a modification of gravity
\cite{BD,DGP,Buc,Ferraris}.

We can parameterize the effective dark energy component in the
modified gravity theories \cite{Linder}. The evolution equation of
Hubble parameter and that of energy density fluctuation can be
simply expressed with this parametrization.

The Palatini formalism where the metric and the connections are
treated as independent variables and the energy momentum tensor
only depends on the metric leads to a different theory from what
is obtained from the metric formalism. The Palatini formalism of
f(R) gravity results in second order differential equations due to
the algebraic relation between the curvature scalar and the trace
of the energy momentum tensor. We will use the evolution of linear
perturbations in f(R) models in the physical frame to check the
stability of the theory. We also investigate the evolution of the
matter density fluctuation.

In the next section we introduce the parametrization of modified
gravities as an effective dark energy. We review the Palatini f(R)
gravity in section 3. In section 4, we investigate the linear
perturbation of f(R) gravity. We also derive the stability
equation of metric fluctuations in the high curvature limit and
show the stability in a specific model. We show the evolutions of
the metric fluctuation and the density fluctuation in the
superhorizon and the subhorizon scales in section 5. We reach our
conclusions in section 6.

\section{Modified Gravity as An Effective Dark Energy}

In general, the modified gravity can be expressed in the following
form \be H^2 - \delta H = \fr{8 \pi G}{3} \rho_{\rm{m}} \, ,
\label{GFri} \ee where $\delta H$ represents the modification to
the Friedmann equation of general relativity and $\rho_{\rm{m}}$
is the matter density. We can rewrite this equation as the
effective dark energy term \be H^2 = \fr{8 \pi G}{3} \Biggl(
\rho_{\rm{m}} + \rho_{\rm{DE}} \Biggr) \, , \label{GFriDE} \ee
where we define the effective dark energy density as $8 \pi G
\rho_{\rm{DE}} = 3 \delta H$. From this equation (\ref{GFriDE}) we
can find the effective equation of state of dark energy \be
\omega_{\rm{DE}} = - 1 - \fr{1}{3} \fr{d \ln \delta H}{d \ln a} \,
. \label{omegaDE} \ee We can also express the expansion history of
the universe in very simple way for the specific case. If $\delta
H$ is proportional to $H$ (i.e. $\delta H = \alpha H$), then we
can find the Hubble parameter $H(z)$ \be \fr{H}{H_0} = \fr{1}{2}
\Biggl( \fr{\alpha}{H_0} + \sqrt{ \Bigl(\fr{\alpha}{H_0} \Bigr)^2
+ 4 \Omega_{\rm{m}}^{(0)} (1 + z)^3} \, \Biggr) \, . \label{H} \ee

\begin{table}[h]
\tbl{Comparison of $\delta H$ and $\omega_{\rm{DE}}$ in BD, DGP,
metric formalism $f(R)$ theory, and Palatini formalism $f(R)$
models. $F(R) = \fr{\partial f(R)}{\partial R}$ and $F_0$ is the
present value of $F(R)$. \label{tab:models}}
{\begin{tabular}{@{}ccc@{}} \toprule & $\delta H$ &
$\omega_{\rm{DE}}$    \\ \colrule BD & $\fr{\omega_{\rm{BD}}}{6}
\fr{\dot{\phi}^2}{\phi^2} - H \fr{\dot{\phi}}{\phi}$ &
$-1+\fr{\fr{\ddot{\phi}}{\phi}-H\fr{\dot{\phi}}{\phi}
+\omega_{\rm{BD}}\fr{\dot{\phi}^2}{\phi^2}}{\fr{\omega_{\rm{BD}}}{2}
\fr{\dot{\phi}^2}{\phi^2}-3H\fr{\dot{\phi}}{\phi}}$ \\
DGP & $\fr{H}{r_0}$ & $-\fr{1}{1 + \Omega_{\rm{m}}}$  \\
Metric $f(R)$& $\fr{1}{3F_0} \Bigl(\fr{1}{2}(FR - f)-3H
\dot{F}+3H^2(F_0 -F) \Bigr)$& $-1 + \fr{2
\ddot{F}-2H\dot{F}-4\dot{H}(F_0-F)}
{FR-f-6H\dot{F}+6H^2(F_0-F)}$ \\
Palatini $f(R)$& $\fr{1}{3F_0} \Bigl(\fr{1}{2}(FR -
f)+\fr{3}{2}\ddot{F}+\fr{3}{2}H \dot{F} -\fr{3}{2}F
\fr{\dot{F}^2}{F^2}+3H^2(F_0 -F) \Bigr)$& $-1+\fr{2\ddot{F}
-2H\dot{F}-3F\fr{\dot{F}^2}{F^2}-4\dot{H}(F_0-F)}{FR-f+3\ddot{F}+3H\dot{F}-
3F \fr{\dot{F}^2}{F^2}+6H^2(F_0-F)}$  \\
\botrule
\end{tabular}}
\end{table}
In table {\ref{tab:models}, we show the expressions of the
modification term to the Friedmann equation ($\delta H$) and the
effective equation of state of dark energy ($\omega_{\rm{DE}}$)
for the different modified gravity models. BD represents
Brans-Dicke theory, DGP stands for the brane model of  Dvali,
Gabadadze, and Porrati \cite{DGP}, metric $f(R)$ means the $f(R)$
theory in the metric formalism \cite{Buc}, and Palatini $f(R)$
does that in the Palatini formalism \cite{Ferraris}. In the
Palatini formalism we can relate the functions $f(R)$ and $F(R)$
to the matter energy density $\rho_{\rm{m}}$.

\begin{table}[h]
\tbl{Comparisons of $H(z)$ and evolution of the linear
perturbation $\delta_{\rm{m}}$ with the effective gravitational
constant $G_{\rm{eff}}$ in each model. In metric and Palatini
formalism the perturbation calculations are held for subhorizon
scale (i.e. $\fr{k}{a} > H$). Where $Q = -\fr{2 F_{,R}}{F}
\fr{k^2}{a^2}$ and $F'(T) = \fr{\partial F(T)}{\partial T}$.
\label{tab:Hmodels}} {\begin{tabular}{@{}cccc@{}} \toprule
& $ \fr{H(z)}{H_0}$ & $\delta_{\rm{m}}$ & $G_{\rm{eff}}$ \\
\colrule
BD & $ \sqrt{\fr{\delta H}{H_0^2} + \fr{\phi_0}{\phi}
\Omega_{\rm{m}}^{(0)}(1+z)^3}$ &$\ddot{\delta} + 2H \dot{\delta} -
4 \pi G_{\rm{eff}} \rho \delta = 0 $ & $
\Biggl(\fr{2\omega_{\rm{BD}}+4}{2\omega_{\rm{BD}}+3}\Biggr) \fr{1}{\phi_0}$ \\
DGP & $\fr{1}{2} \Biggl( \fr{1}{r_0H_0} + \sqrt{
\Bigl(\fr{1}{r_0H_0} \Bigr)^2 + 4 \Omega_{\rm{m}}^{(0)} (1 + z)^3}
\, \Biggr)$ & $\ddot{\delta} + 2H \dot{\delta} - 4 \pi
G_{\rm{eff}} \rho \delta = 0 $ &
$G \Bigl(1 + \fr{1}{3[1 + 2 r_0 H \omega_{\rm{DE}}]} \Bigr)$ \\
Metric $f(R)$& $\sqrt{\fr{\delta H}{H_0^2} + \Omega_{\rm{m}}^{(0)}
( 1 + z)^3}$ & $\ddot{\delta} + 2H \dot{\delta} - 4 \pi
G_{\rm{eff}} \rho \delta = 0 $
& $\fr{2(1-2Q)}{2-3Q}\fr{G}{F}$ \\
Palatini $f(R)$ & $\sqrt{\fr{\delta H}{H_0^2} +
\Omega_{\rm{m}}^{(0)} ( 1 + z)^3}$ &
$\ddot{\delta}+H\dot{\delta}-4\pi G_{\rm{eff}}\rho\delta=0$&
$-\fr{1}{4\pi}\fr{F'}{2F+3F'\rho}\fr{k^2}{a^2}$ \\
\botrule
\end{tabular}}
\end{table}

We show the expansion rate $H(z)/H_0$ where $H_0$ is the present
value of the Hubble parameter, the evolution equation for the
energy density perturbation $\delta_m$, and the effective
gravitational constant $G_{\rm{eff}}$ for different models in
table \ref{tab:Hmodels}. The evolution equations of $\delta_m$ for
the $f(R)$ models are suitable for the subhorizon scale. The scale
dependence on $G_{\rm{eff}}$ is the one of the major differences
compared to the general relativity.

\section{Palatini f(R) Gravity}

We consider a modification to the Einstein-Hilbert action assuming
the Palatini formalism, where the metric $g_{\mu\nu}$ and the
torsionless connection $\hat{\Gamma}^{\alpha}_{\mu\nu}$ are
independent quantities and the matter action depends only on
metric \be S = \int d^4 x \sqrt{-g} \Biggl[ \fr{1}{2 \kappa^2} f
\Bigl(\hat{R}(g_{\mu\nu}, \hat{\Gamma}^{\alpha}_{\mu\nu}) \Bigr) +
{\cal L}_{m}(g_{\mu\nu}, \psi) \Biggr] \, , \label{action} \ee
where $\psi$ are matter fields. Then the Ricci tensor is defined
solely by the connection \be \hat{R}_{\mu\nu} =
\hat{\Gamma}^{\alpha}_{\mu\nu , \,\alpha} -
\hat{\Gamma}^{\alpha}_{\mu\alpha  , \,\nu} +
\hat{\Gamma}^{\alpha}_{\alpha\beta}\hat{\Gamma}^{\beta}_{\mu\nu} -
\hat{\Gamma}^{\alpha}_{\mu\beta}\hat{\Gamma}^{\beta}_{\alpha\nu}
\, , \label{hatRmunu} \ee whereas the scalar curvature is given by
$\hat{R} = g^{\mu\nu} \hat{R}_{\mu\nu}$. Then we can find the
Ricci tensor and the scalar curvature by using the metric relation
\ba \hat{R}_{\mu\nu} &=& R_{\mu\nu} + \fr{3}{2} \fr{1}{F^2}
\nabla_{\mu} F \nabla_{\nu} F - \fr{1}{F} \nabla_{\mu}
\nabla_{\nu} F - \fr{1}{2} g_{\mu\nu} \fr{1}{F} \Box F \, ,
\label{hatRmunu2} \\ \hat{R} &=& R - 3 \fr{1}{F} \Box F +
\fr{3}{2} \fr{1}{F^2} (\partial F)^2 \, . \label{hatR2} \ea We can
derive the field equation of f(R) gravity in the Palatini
formalism from the above action (\ref{action}) \be F(\hat{R})
\hat{R}_{\mu\nu} - \fr{1}{2} g_{\mu\nu} f(\hat{R}) = 8\pi G
T_{\mu\nu} \, , \label{seq} \ee where $F(\hat{R}) = \partial
f(\hat{R}) /
\partial \hat{R}$ and the matter energy momentum tensor is given
as usual form $T_{\mu\nu} = - \fr{2}{\sqrt{-g}}
\fr{\delta(\sqrt{-g}{\cal L}_m)}{\delta g^{\mu\nu}}$. From the
above equations we can get the generalized Einstein equation \be
G_{\mu\nu} = 8\pi G T_{\mu\nu} + ( 1 - F) R_{\mu\nu} - \fr{3}{2}
\fr{1}{F} \nabla_{\mu} F \nabla_{\nu} F + \nabla_{\mu}
\nabla_{\nu} F + \fr{1}{2} (f - R) g_{\mu\nu} + \fr{1}{2}
g_{\mu\nu} \Box F \, . \label{Gmunu} \ee From the above equation
we can derive the modified Friedmann equations \ba 3 H^2 &=&
\fr{8\pi G}{2F} \rho + \fr{f}{2F} - 3H \fr{\dot{F}}{F} - \fr{3}{4}
\fr{\dot{F}^2}{F^2} \, , \label{G001} \\
-2 \dot{H} &=& \fr{8 \pi G}{F} (\rho + p) +  \fr{\ddot{F}}{F} -
H\fr{\dot{F}}{F} - \fr{3}{2} \fr{\dot{F}^2}{F^2} \, , \label{G002}
\ea where dot denote derivatives with respect to $t$.

\section{Linear Perturbation and Stability in Palatini f(R) Gravity}

The line element in the conformal Newtonian gauage is given by \be
ds^2 = a^2(\tau) \Biggl[ -\Bigl(1 + 2 \Psi(\tau, \vec{x}) \Bigr)
d\tau^2 + \Bigl(1 - 2 \Phi(\tau,\vec{x}) \Bigr) dx^i dx_i \Biggr]
\, . \label{CNG} \ee The main modifications for viable models with
stable high curvature limits happen well during the matter
dominated epoch and we can take the components of the energy
momentum tensor as \be T^{0}_{0} = -\rho (1 + \delta) \, ,
\hspace{0.2in} T^{0}_{i} = \rho \partial_i q \, , \hspace{0.2in}
T^{i}_{j} = 0 \, . \label{Tmunumatter} \ee From the equation
(\ref{Gmunu}), we can find the perturbed Einstein equation \ba F
\delta G^{\mu}_{\nu} &=& \kappa^2 \delta T^{\mu}_{\nu} -
R^{\mu}_{\nu} \delta F - \fr{3}{2} \fr{\delta(\nabla^{\mu} F
\nabla_{\nu} F)}{F} + \fr{3}{2} \fr{\nabla^{\mu} F \nabla_{\nu}
F}{F^2} \delta F + \delta(\nabla^{\mu} \nabla_{\nu} F) \nonumber
\\ && + \Biggl( \fr{3}{2} \fr{\Box F}{F} \delta F + \fr{3}{4}
\fr{\delta (\partial F)^2}{F} - \fr{3}{2} \fr{(\partial F)^2}{F^2}
\delta F - \delta \Box F \Biggr) \delta^{\mu}_{\nu} \, ,
\label{deltaGmunu} \ea where we use $\delta f(\hat{R}) =
F(\hat{R}) \delta \hat{R}$. If we consider the $ij$-component of
the perturbed equation, then we can find \be \Phi - \Psi =
\fr{\delta F}{F} \, , \label{deltaGij} \ee where we assume the
null anisotropic stress. If we use the above equation
(\ref{deltaGij}), then we can express the other components of the
perturbed Einstein equation (\ref{deltaGmunu}) \ba && 3H^2 \Biggl[
\Phi' + \Psi' + \fr{1}{2} \fr{F'}{F} (\Phi' + \Psi') + \Bigl(
\fr{1}{2} \fr{F''}{F} - \fr{1}{2} \fr{F'^2}{F^2} + \fr{1}{2}
\fr{H'}{H} \fr{F'}{F}  + \fr{1}{2} \fr{F'}{F} + \fr{H'}{H} + 1
\Bigr) \Phi \nonumber \\ && + \Bigl( - \fr{1}{2} \fr{F''}{F} +
\fr{F'^2}{F^2} - \fr{1}{2} \fr{H'}{H} \fr{F'}{F} + \fr{3}{2}
\fr{F'}{F} - \fr{H'}{H} + 1 \Bigr) \Psi
 \Biggr] + \fr{k^2}{a^2} (\Phi + \Psi) =
 - \fr{\kappa^2 \rho}{F} \delta \, , \label{deltaG00c1}
\\ && H \Biggl[\Phi' + \Psi' + \Phi + \Psi + \fr{1}{2} \fr{F'}{F}
(\Phi + \Psi) \Biggr] = - \fr{\kappa^2 \rho}{F} q \, ,
\label{deltaG0ic1} \\ && 3 H^2 \Biggl[ \Phi'' + \Psi'' + \Bigl( 4
+ \fr{H'}{H} \Bigr) \Phi' + \Bigl(3 \fr{F'}{F} + 4 + \fr{H'}{H}
\Bigr) \Psi' + \Bigl(- \fr{F''}{F} - (2 + \fr{H'}{H}) \fr{F'}{F}
\nonumber \\ && + \fr{H'}{H} + 3 \Bigr) \Phi + \Bigl( 3
\fr{F''}{F} + (6 + 3 \fr{H'}{H}) \fr{F'}{F} + 3 \fr{H'}{H} + 3
\Bigr) \Psi \Biggr] = 0 \, . \label{deltaGiic1} \ea where primes
denote derivatives with respect to $\ln a$. To capture the metric
evolution, let us introduce two parameters as in the reference
\cite{Hu}: $\theta$ the deviation from $\zeta$ conservation and
$\epsilon$ the deviation from the superhorizon metric evolution
\ba \zeta' = \Phi' + \Psi - H' q &=& - \fr{H'}{H}
\Biggl(\fr{k}{aH} \Biggr)^2 B \theta \, , \label{theta} \\ \Phi''
+ \Psi' - \fr{H''}{H'} \Phi' + \Biggl(\fr{H'}{H} - \fr{H''}{H'}
\Biggr) \Psi &=& - \Biggl(\fr{k}{aH} \Biggr)^2 B \epsilon  \, ,
\label{epsilon} \ea where we define the dimensionless quantity \be
B = \fr{F'}{F} \fr{H}{H'} \, . \label{B} \ee From the above
equations, we can find the expression for $\theta$ and $\epsilon$,
\ba \fr{H'}{H} \Biggl(\fr{k}{aH} \Biggr)^2 B \theta &=& \fr{1}{2}
\Biggl[ \fr{B'}{B} + \fr{3}{2} \fr{H'}{H} B + \fr{H''}{H'} + 4
\Biggr] (\Phi - \Psi) + \fr{1}{2} \Biggl[ -\fr{H'}{H} B' + \Bigl(
\fr{H'}{H} - \fr{H''}{H} \Bigr) B \nonumber \\ && + \fr{1}{2}
\fr{H'^2}{H^2} B^2 \Biggr] H q
\label{theta2} \\
\Biggl(\fr{k}{aH} \Biggr)^2 B
 \epsilon &=& - \fr{1}{2}
 \Biggl[ 2 \fr{B'^2}{B^2} + \Bigl( 5 \fr{H''}{H'} + \fr{H'}{H} + 9 \Bigr) \fr{B'}{B}
 + \fr{H'}{H} B' + \fr{H'^2}{H^2} B^2 + \Bigl( \fr{5}{2}
 \fr{H''}{H} + \fr{3}{2} \fr{H'^2}{H^2} \nonumber \\ && + 6 \fr{H'}{H} \Bigr) B +
 \fr{H''}{H} + 3 \fr{H''^2}{H'^2} + 13 \fr{H''}{H'} + \fr{H'}{H} +
 9 + 2 \Bigl( \fr{H''}{H'} + \fr{H'}{H} + 3 \Bigr) \fr{1}{B}
 \Biggr] \nonumber \\ && (\Phi - \Psi)  + \Biggl[ \fr{H'}{H} B'
 + \fr{1}{2} \fr{H'^2}{H^2} B^2 +
 \Bigl( \fr{H''}{H} + \fr{3}{2} \fr{H'}{H} \Bigr) B \Biggr] \Psi - \fr{1}{2}
 \Biggl[ \fr{H'}{H} B + \fr{H''}{H'} \nonumber \\ && + \fr{H'}{H} + 3 \Biggr]
 \fr{\kappa^2 \rho}{F H} q  \label{epsilon2}
 \ea From equation (\ref{theta2}), we can recover the conservation
 of Newtonian gauge when $\Phi = \Psi$ and $F$ is a constant.

 In addition to these equations, we can find very useful equation
 from the structure equation (\ref{seq})
 \be \fr{\delta F}{F} = - \fr{1}{3} \fr{F'}{F} \delta = \Phi -
 \Psi \, . \label{deltaF} \ee From this equation we can find the
 evolution equation of matter density fluctuation \ba && \delta'' + \Biggl( 2
 \fr{B'}{B} + \fr{H'}{H} B + 2 \fr{H''}{H'} - \fr{H'}{H} + 3
 \Biggr) \delta' + \Biggl( \fr{H'}{H} B' - 2 \fr{B'^2}{B^2} -
 \Bigl[ 4 \fr{H''}{H'} + 2 \fr{H'}{H} + 4 \Bigr] \fr{B'}{B}
 \nonumber \\ && +
 \fr{H'^2}{H^2} B^2 + \Bigl[ \fr{H''}{H} - \fr{H'^2}{H^2} + 5
 \fr{H'}{H} \Bigr] B + \Bigl[ - 3 \fr{H''}{H} - 4 \fr{H''}{H'} -
 2 \fr{H''^2}{H'^2} + \fr{H'^2}{H^2} - 2 \fr{H'}{H} + 6 \Bigr]
 \nonumber \\ && - 4
 \Bigl[\fr{H''}{H'} + \fr{H'}{H} + 3 \Bigr] \fr{1}{B} \Biggr)
 \delta = - 3 \Psi' \, . \label{deltadoubleprime} \ea Compared
 with previous works \cite{perturb}, we do not specify the gauge
 of matter density to solve the matter density fluctuation.

 Unstable metric fluctuations can create order unity effects that
invalidate the background expansion history. We can derive the
evolution equation of the deviation parameter \cite{SLee}. If we
differentiate the equation (\ref{epsilon2}) and consider the
evolution in the superhorizon scale, then we have \ba &&
\epsilon'' + \Biggl( 2 \fr{B'}{B} + \fr{H'}{H} B + \fr{H''}{H'} -
3 \fr{H'}{H} - 1 \Biggr) \epsilon' + \Biggl( - 2 \fr{B'^2}{B^2} +
2 \fr{H'}{H} B' - \Bigl[ 5 \fr{H''}{H'} + 4 \fr{H'}{H} + 9 \Bigr]
\fr{B'}{B} \nonumber \\ && + \fr{1}{2} \fr{H'^2}{H^2} B^2 + \Bigl[
2 \fr{H''}{H'} - 4 \fr{H'^2}{H^2} + \fr{3}{2} \fr{H'}{H} \Bigr] B
+ Q' +  \Bigl[ -2 \fr{H''}{H'} - \fr{H'}{H} + 1 \Bigr] Q + 4
\fr{H'^2}{H^2} + 6 \fr{H'}{H} + 7 \nonumber \\
&& - 4 \fr{Q}{B} \Biggr) \epsilon = \fr{1}{B} F(\Psi, \Phi, Hq) \,
, \label{epsilonevol} \ea where we use equations
(\ref{deltaGiic1}) and (\ref{epsilon}) and $F(\Psi, \Phi)$ is the
source function for the deviation $\epsilon$ and define $Q$ as \be
Q = \fr{H''}{H'} + \fr{H'}{H} + 3 \label{Q} \, . \ee The above
equation is different from that of the metric formalism \cite{Hu}.
The stability of $\epsilon$ depends on the sign of the coefficient
of the term proportional to $\epsilon$. In the metric formalism
$\epsilon$ is stable as long as $B > 0$. However, the stability is
complicate and need to be checked for each model in the Palatini
formalism.

\subsection{A particular example : $ f(\hat{R}) = \beta \hat{R}^{n}$}

We demonstrate the general consideration of the previous
subsection with a specific choice for the nonlinear Lagrangian,
$f(\hat{R}) = \beta \hat{R}^n$, where $n \neq 0, 2, 3$. The
background is simply described by a constant effective equation of
state in this model. The Hubble parameter scales as $H^2 \sim
a^{-3/n}$. Then it is easy to write it with its derivatives in
terms of $\ln a$ \be \fr{H'}{H} = - \fr{3}{2n} \, , \hspace{0.2in}
\fr{H''}{H} = \Biggl( - \fr{3}{2n} \Biggr)^2 \, . \label{H2} \ee
Here the scalar curvature is $\hat{R} = 3(3 - n) H^2 / (2n)$. From
this fact, we can also find the derivatives of $F$ with respect to
$\ln a$ \be \fr{F'}{F} = \fr{F''}{F'} = \fr{3(1 - n)}{n} \, ,
\hspace{0.2in} \fr{F''}{F} = \fr{F'''}{F'} = \Biggl( \fr{3(1 -
n)}{n} \Biggr)^2 \, . \label{F2} \ee If we use above equations
(\ref{H2}) and (\ref{F2}) into (\ref{epsilon2}), then we find that
the deviation from the superhorizon metric evolution is null,
$\epsilon = 0$.

\section{Evolutions of Metric and Matter Density}

\subsection{Superhorizon evolution}

We consider the metric evolution in superhorizon sized, $k/(aH)
\ll 1$. In this case, the anisotropy relation of the equation
(\ref{theta2}) becomes \be \Phi - \Psi \simeq \Bigl( B + A \Bigr)
H'q \, , \label{superPhi} \ee where $A$ is given by \be A = -
\fr{B \Bigl( 2 B \fr{H'}{H} + 5 \Bigr)}{\Bigl(\fr{B'}{B} +
\fr{3}{2} \fr{H'}{H} B + \fr{H''}{H'} + 4 \Bigr)} \, . \label{A}
\ee From the above equations we can find the superhorizon
evolution equation of $\Phi$ and $\delta$ \ba && \Phi'' + \Biggl(
\fr{B'}{B} + 2 \fr{H'}{H} B + \fr{H''}{H'} - \fr{H'}{H} + 4 - C
\Biggr) \Phi' + \Biggl( \fr{B'}{B} + \fr{H'}{H} B + \fr{H''}{H'} +
3 - C \Biggr) \Phi \simeq 0 \, , \label{superPhidoubleprime} \nonumber \\
&& \Biggl( 1 + \fr{B}{B + A} \Biggr) \delta'' + \Biggl( 2
 \fr{B'}{B} + \fr{H'}{H} B + 2 \fr{B'A - BA'}{(B + A)^2}
 - \fr{H'}{H} \fr{B}{B + A} + 2 \fr{H''}{H'} - \fr{H'}{H} + 3
 \Biggr) \delta' \nonumber \\ && + \Biggl( \fr{H'}{H} B'
 - 2 \fr{B'^2}{B^2} -
 \Bigl[ 4 \fr{H''}{H'} + 2 \fr{H'}{H} + 4 \Bigr] \fr{B'}{B}
 + \fr{H'^2}{H^2} B^2 + \Bigl[ \fr{H''}{H} - \fr{H'^2}{H^2} + 5
 \fr{H'}{H} \Bigr] B \nonumber \\ && \fr{B''A - BA''}{(B + A)^2}
 - 2 \fr{(B'A -
BA')}{(B + A)^2} \fr{(B' + A')}{(B + A)}  - \fr{H'}{H} \fr{B'A -
BA'}{(B + A)^2} - \Biggl(\fr{H'}{H} \Biggr)' \fr{B}{B + A}
\nonumber \\ && + \Bigl[ - 3 \fr{H''}{H} - 4 \fr{H''}{H'} -
 2 \fr{H''^2}{H'^2} + \fr{H'^2}{H^2} - 2 \fr{H'}{H} + 6 \Bigr] - 4
 \Bigl[\fr{H''}{H'} + \fr{H'}{H} + 3 \Bigr] \fr{1}{B} \Biggr)
 \delta = 0  \label{superdeltadoubleprime} \, ,
\ea where $C$ is defined as \be C = \fr{1}{B + A + 1} \Biggl[
\fr{B'}{B} + 2 \fr{H'}{H} B + 2 \fr{H''}{H'} - \fr{H'}{H} + 3
\Biggr] \, . \label{C} \ee

\subsection{Superhorizon evolution in a particular example}

Now we can check the evolution equations in the previous
subsection in a particular case, $f(\hat{R}) \sim \hat{R}^{n}$. In
this case, we can simplify the following quantities \be B = 2 (n -
1) \, , \hspace{0.2in} A = - 4 (n - 1) = - 2B \, , \hspace{0.2in}
C = \fr{3}{2n} = - \fr{H'}{H} \, . \label{BAC} \ee From this, we
can also simplify the evolution equations
(\ref{superPhidoubleprime}) and (\ref{superdeltadoubleprime}) \ba
\Phi'' + \fr{9 - 4n}{2n} \Phi' &=& 0 \, , \label{sPdp2} \\ \delta'
&=& 0 \, . \label{sddp2} \ea The Newtonian potential $\Phi =$
constant is a solution to the equation. Also the matter density
fluctuation has the same for as general relativity, $\delta =$
constant.

\subsection{Subhorizon evolution}

For subhorizon scales where $k/aH \gg 1$,we can find the Poisson
equation from the equation (\ref{deltaG00c1}) \be k^2 (\Phi +
\Psi) \simeq - \fr{\kappa^2 a^2 \rho}{F} \delta \, .
\label{Poisson} \ee If we use equations (\ref{deltaF}) and
(\ref{Poisson}), then we have \be 3 \Psi \simeq \Biggl( - 3
\fr{\kappa^2 \rho}{F H^2} \fr{a^2 H^2}{k^2} + \fr{F'}{F} \Biggr)
\fr{\delta}{2} \simeq \fr{F'}{F} \fr{\delta}{2} \simeq - 3 \Phi \,
. \label{subPsi} \ee We can find the evolution equations of $\Phi$
and $\delta$ \ba && \Phi'' + \Biggl(\fr{B'}{B} + \fr{5}{2}
\fr{H'}{H} B + \fr{H''}{H'} + \fr{H'}{H} + 6 \Biggr) \Phi' -
\Biggl( \fr{B'^2}{B^2} + \Bigl[2
\fr{H''}{H'} - 1 \Bigr] \fr{B'}{B} - 2 \fr{H'}{H} B' \nonumber \\
&& - \fr{9}{4} \fr{H'^2}{H^2} B^2  + \Bigl[ 2 \fr{H''}{H} - 4
\fr{H''}{H'} - \fr{21}{2} \fr{H'}{H} \Bigr] B + \Bigl[
\fr{H''^2}{H'^2} - \fr{H''}{H'} + 10 \fr{H'}{H} + 6 \Bigr]
\nonumber \\ && + \Bigl[2 \fr{H''}{H'} + 2 \fr{H'}{H} + 6 \Bigr]
\fr{1}{B} \Biggr) \Phi \simeq 0 \, . \label{subPhidoubleprime} \\
&& \delta'' + \Biggl( 2
 \fr{B'}{B} + \fr{3}{2} \fr{H'}{H} B + 2 \fr{H''}{H'} - \fr{H'}{H} + 3
 \Biggr) \delta' + \Biggl( \fr{3}{2} \fr{H'}{H} B' - 2 \fr{B'^2}{B^2} -
 \Bigl[ 4 \fr{H''}{H'} + 2 \fr{H'}{H} + 4 \Bigr] \fr{B'}{B}
 \nonumber \\ && +
 \fr{H'^2}{H^2} B^2 + \Bigl[ \fr{3}{2} \fr{H''}{H} - \fr{3}{2} \fr{H'^2}{H^2}
 + 5 \fr{H'}{H} \Bigr] B + \Bigl[ - 3 \fr{H''}{H} - 4 \fr{H''}{H'} -
 2 \fr{H''^2}{H'^2} + \fr{H'^2}{H^2} - 2 \fr{H'}{H} + 6 \Bigr]
 \nonumber \\ && - 4
 \Bigl[\fr{H''}{H'} + \fr{H'}{H} + 3 \Bigr] \fr{1}{B} \Biggr)
 \delta = 0 \, . \label{subdeltadoubleprime} \ea

\subsection{Subhorizon evolution in a particular example}

We can use the previous relation (\ref{BAC}) into the evolution
equations (\ref{subPhidoubleprime}) and
(\ref{subdeltadoubleprime}) \ba \Phi'' + \fr{3(3 - n)}{2n} \Phi'
+ \fr{3(14n^2 + 19n -36)}{4 n^2} \Phi &=& 0 \label{subPdp} \\
\delta'' + \fr{3(2 - n)}{2n} \delta' &=& 0 \label{subddp} \ea The
subhorizon scale evolutions of $\Phi$ and $\delta$ show different
behaviors from those of general relativity as expected
\cite{PZhang}.

\section{Conclusions}

We investigate the different modified gravity models as an
effective dark energy models. We can parameterize the effective
equation of state of dark energy in terms of modified gravities.
We show the evolution of energy density fluctuation compared to
that of general relativity.

We have analyzed the stability of metric fluctuations by checking
the cosmological evolution of linear perturbations in Palatini
f(R) gravity. We have also considered the matter density
fluctuation in the Newtonian gauge.

We have shown that the stability of metric fluctuations in the
high redshift limit of high curvature is not simply expressed. We
need to check each model for the stability. However, we have found
that the deviation from the superhorizon metric evolution is null
for a specific choice of the nonlinear Einstein-Hilbert action,
$f(\hat{R}) \sim \hat{R}^{n}$ and stability of this model is
guaranteed.

We have investigated the evolution equations of Newtonian
potential and matter density contrast in super and sub-horizon
scales. In the specific model, superhorizon evolutions of
Newtonian potential and matter density fluctuation are same to
those of general relativity. However, subhorizon evolutions show
the different behaviors from the general relativity case. This
will give us the method to probe the possibility of $f(R)$ theory.

\section*{Acknowledgments}

We thank CosPA2007 organizing committee for their hospitality and
for organizing such a nice meeting.

\end{document}